\documentclass[prd,preprint,tightenlines,floatfix,showpacs,preprintnumbers,nofootinbib,eqsecnum,superscriptaddress]{revtex4}

 \usepackage[dvips,final]{graphicx}
  \usepackage{amssymb}
    \usepackage{amsfonts}
     \usepackage{epsfig}
      \usepackage{bm}

\def\Pom{{\bf I\!P}}

\newcommand{\bDelta}{\mbox{\boldmath $\Delta$}}

\newcommand{\bkappa}{\mbox{\boldmath $\kappa$}}
\newcommand{\bp}{\mbox{\boldmath $p$}}

\newcommand{\bk}{\mbox{\boldmath $k$}}

\newcommand{\bM}{\mbox{\boldmath $M$}}

\newcommand{\half}{{1\over 2}}
\def\lsim{\mathrel{\rlap{\lower4pt\hbox{\hskip1pt$\sim$}}
    \raise1pt\hbox{$<$}}}         
\def\gsim{\mathrel{\rlap{\lower4pt\hbox{\hskip1pt$\sim$}}
    \raise1pt\hbox{$>$}}}         

\begin{document}

\thispagestyle{empty} \preprint{\hbox{}} \vspace*{-10mm}

\title{Exclusive photoproduction of $\Upsilon$:  
from HERA to Tevatron}

\author{A. Rybarska}
\email{Anna.Cisek@ifj.edu.pl}
\affiliation{Institute of Nuclear Physics PAN, PL-31-342 Cracow,
Poland} 
\author{W. Sch\"afer}
\email{Wolfgang.Schafer@ifj.edu.pl}
\affiliation{Institute of Nuclear Physics PAN, PL-31-342 Cracow,
Poland} 
\author{A. Szczurek}
\email{Antoni.Szczurek@ifj.edu.pl}
\affiliation{Institute of Nuclear Physics PAN, PL-31-342 Cracow,
Poland} 
\affiliation{University of Rzesz\'ow, PL-35-959 Rzesz\'ow,
Poland}

\date{\today}

\begin{abstract}
The forward photoproduction amplitude for $\gamma p \to \Upsilon p$
is calculated in a pQCD $k_\perp$-factorization approach with
an unintegrated gluon distribution constrained 
by inclusive deep--inelastic structure functions.
The total cross section for diffractive $\Upsilon$s is 
compared with a recent HERA data. We also discuss the $2S/1S$ ratio
in diffractive $\Upsilon$--production.
The amplitude is used to predict the cross section for exclusive
$\Upsilon$ production in hadronic reactions.
Differential distributions for the exclusive 
$p \bar p \to p \Upsilon \bar p$ process
are calculated for Tevatron energies. We also show predictions for LHC.
Absorption effects are included. 
\end{abstract}


\maketitle

\section{Introduction}

The inclusive production of quarkonia was studied intensively in the past
both in elementary hadronic and nuclear reactions at SPS, RHIC and Tevatron
energies. For a review see e.g.\cite{quarkonia_reviews}.
In contrast, the exclusive production of heavy $Q\bar Q$ vector
quarkonium states (e.g. $h_1 h_2 \to h_1 \Upsilon h_2$) in hadronic 
interactions was never measured, but attracted recently 
much attention from the theoretical side \cite{SMN,KMR,Klein,GM05,Bzdak,SS07,GM}. 
Due to the negative charge-parity of the vector meson, 
the purely hadronic Pomeron--Pomeron fusion 
mechanism of exclusive meson production is not available, 
and instead the production will proceed via photon--Pomeron fusion. 
A possible purely hadronic mechanism would involve the elusive Odderon exchange
\cite{SMN,Bzdak}. Currently there is no compelling evidence
for the Odderon, and here we restrict ourselves to the photon--exchange
mechanism, which exists without doubt, and must furthermore
dominate any hadronic exchange at very small momentum transfers.
In our approach to the exclusive hadronic reaction, we follow
closely the procedure outlined in our previous work on $J/\psi$
production \cite{SS07}. There is one crucial difference, though.
While in the case of diffractive $J/\psi$ photoproduction there 
exist a large body of fairly detailed data, including e.g. transverse
momentum distributions, the photoproduction data for exclusive
$\Upsilon$'s are rather sparse \cite{ZEUS_old,H1,ZEUS_Upsilon}. 
Hence, different from \cite{SS07} we cannot avoid modelling the 
relevant $\gamma p \to \Upsilon p$ subprocess.
Fortunately, due to the large mass of the $\Upsilon$'s constituents,
the cross section gets its main contribution from 
small--size $b \bar b$--dipoles, and the production mechanism 
can be described in a pQCD framework (for a recent review and references,
see \cite{INS06}). 
The two main ingredients are the unintegrated gluon distribution
of the proton, and the light--cone wave function of the vector
meson. The unintegrated gluon distribution is sufficiently 
well constrained by the precise small--$x$ data for the 
inclusive proton structure function,
and we shall content ourselves here with a particular parametrisation
which provides a good description of inclusive deep inelastic scattering 
data \cite{IN_glue}. As the relevant energy range of the 
$ \gamma p \to \Upsilon p$ subprocess at Tevatron overlaps well with the 
HERA energy range, any glue which fulfills the stringent constraints of the precise
HERA $F_2$--data must do a similar job. Alternative unintegrated
gluon distributions are discussed for example in \cite{Unintegrated}.

The current experimental analyses at the Tevatron \cite{Pinfold} 
call for an evaluation of differential distributions 
including the effects of absorptive corrections.

The HERA data cover the $\gamma p$ center of mass (cm--) 
energy range $W \sim 100 \div 200$ GeV. 
This energy range is in fact very much relevant to the exclusive production
at Tevatron energies for not too large rapidities of the meson, say $|y| \lsim 3$.
This will be different at the LHC, where 
a broad range of subsystem energies $W_{\gamma p}$, up to several TeV, 
is spanned for $\Upsilon$ emitted in the forward directions.
This will require a long-range extrapolation to a 
completely new unexplored region.
In this paper, however, we will concentrate on predictitions for Tevatron energies.
Here our input amplitude is constrained by the HERA data, to which description
we now turn.

\section{Photoproduction $\gamma p \to \Upsilon p$ at HERA}

We thus turn to the analysis of the photoproduction recation
studied at HERA. The photoproduction amplitude will then be the major
building block for our prediction of exclusive $\Upsilon$ production
at the Tevatron.

\subsection{Amplitude for $\gamma p \to \Upsilon p$}

\begin{figure}[!h]    %
\includegraphics[width=0.4\textwidth]{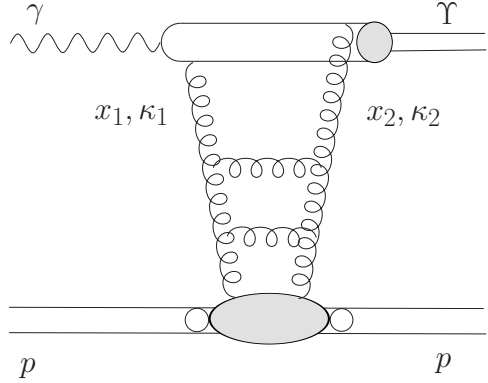}
   \caption{\label{fig:diagram_photon_pomeron}
   \small  A sketch of the exclusive $\gamma p \to \Upsilon p$ amplitude.}
\end{figure}

The amplitude for the reaction under consideration
is shown schematically in Fig.\ref{fig:diagram_photon_pomeron}.
As it is explained in Ref.\cite{INS06}, 
the imaginary part of the amplitude for the $\gamma^* p \to \Upsilon p$
process can be written as
\begin{eqnarray}
\Im m \; {\cal M}_{\lambda_{\gamma},\lambda_V}(W,t=-\bDelta^2,Q^2) =
W^2 \frac{c_\Upsilon \sqrt{4 \pi \alpha_{em}}}{4 \pi^2}
\int \frac{d^2\bkappa}{\kappa^4} \alpha_S(q^2)  
{\cal F}(x_1,x_2,\bkappa_1,\bkappa_2) 
\nonumber \\
\times
\int \frac{dz d^2 \bk}{z (1-z)}  
I_{\lambda_{\gamma}, \lambda_V}(z,\bk,\bkappa_1,\bkappa_2,Q^2) \; ,
\label{full_imaginary}
\end{eqnarray}
where the transverse momenta of gluons coupled to the $Q \bar Q$
pair can be written as
\begin{eqnarray}
\bkappa_1 = \bkappa + {\bDelta \over 2} \, , \, 
\bkappa_2 = - \bkappa + {\bDelta \over 2} \, .
\end{eqnarray}
The quantity ${\cal F}(x_1,x_2,\bkappa_1,\bkappa_2)$ is
the off diagonal unintegrated gluon distribution. 
Explicit expressions for $I_{\lambda_{\gamma}, \lambda_V}$ 
can be found in \cite{INS06}.
For heavy vector mesons, helicity--flip transitions may be neglected,
and we concentrate on the $s$--channel helicity conserving amplitude,
$\lambda_\gamma = \lambda_V$. In the forward scattering limit, i.e.
for $\bDelta =0$, azimuthal integrations can be performed analytically,
and we obtain the following representation for the
imaginary part of the amplitude for forward photoproduction $\gamma p \to \Upsilon p$ :
\begin{eqnarray}
\Im m \, {\cal M}(W,\Delta^2 = 0,Q^2=0) =
W^2 \frac{c_\Upsilon \sqrt{4 \pi \alpha_{em}}}{4 \pi^2} \, 2 \, 
 \int_0^1 \frac{dz}{z(1-z)}
\int_0^\infty \pi dk^2 \psi_V(z,k^2) \\
\int_0^\infty
 {\pi d\kappa^2 \over \kappa^4} \alpha_S(q^2) {\cal{F}}(x_{eff},\kappa^2)
\Big( A_0(z,k^2) \; W_0(k^2,\kappa^2) 
     + A_1(z,k^2) \; W_1(k^2,\kappa^2)
\Big) \, ,
\label{amplitude_forward}
\end{eqnarray}
where
\begin{eqnarray}
A_0(z,k^2) &=& m_b^2 + \frac{k^2 m_b}{M + 2 m_b}  \, , \\
A_1(z,k^2) &=& \Big[ z^2 + (1-z)^2 
    - (2z-1)^2 \frac{m_b}{M + 2 m_b} \Big] \, \frac{k^2}{k^2+m_b^2} \, ,
\end{eqnarray}
and
\begin{eqnarray}
W_0(k^2,\kappa^2) &=& 
{1 \over k^2 + m_b^2} - {1 \over \sqrt{(k^2-m_b^2-\kappa^2)^2 + 4 m_b^2 k^2}}
\, , 
\nonumber \\
W_1(k^2,\kappa^2) &=& 1 - { k^2 + m_b^2 \over 2 k^2}
\Big( 1 + {k^2 - m_b^2 - \kappa^2 \over 
\sqrt{(k^2 - m_b^2 - \kappa^2)^2 + 4 m_b^2 k^2 }}
\Big) \, .
\end{eqnarray}

To obtain these results, the perturbative $\gamma \to b \bar b$ light cone wave
function was used; the vertex for the $b \bar b \to \Upsilon$  transition
is given below, and is obtained by projecting onto the pure $s$--wave 
$b \bar b$--state.
Here $c_\Upsilon = e_b = -1/3$, and the mass of the bottom quark is taken as
$m_b = 4.75$ GeV. 
The relative transverse momentum squared of (anti-)quarks in the bound state
is denoted by $k^2$, their longitudinal momentum fractions are $z,1-z$, and
we introduced
\begin{equation}
M^2 = {k^2 + m_b^2 \over z(1-z)} \, .
\end{equation}
The dominant contribution to the amplitude comes from the piece $
\propto A_0 W_0 \sim m_b^2 W_0$, and our exact projection onto 
$s$--wave states differs in fact only marginally from the 
naive $\gamma_\mu$--vertex for the $\Upsilon \to b \bar b$ transition.
The 'radial' light-cone wave function of the vector meson,
$\psi_V(z,k^2)$ will be discussed further below.
The unintegrated gluon distribution ${\cal{F}}(x,\kappa^2)$ is normalized such that
for a large scale $\bar Q^2$ it will be related to the integrated gluon
distribution $g (x,\bar Q^2)$ through
\begin{equation}
x g(x,\bar Q^2 ) = \int^{\bar Q^2} {d\kappa^2 \over \kappa^2} {\cal{F}}(x,\kappa^2) \, . 
\end{equation}
The running coupling $\alpha_S$ enters at the largest relevant virtuality 
$q^2 = \max \{ \kappa^2, k^2 + m_b^2 \}$.
Due to the finite mass of the final state vector meson, the 
longitudinal momentum transfer is nonvanishing, and, as indicated above,
a more precise treatment would require the use of skewed/off--diagonal 
gluon distributions.
At the high energies relevant here it is admissible to account for skewedness
by an appropriate rescaling of the diagonal gluon distribution \cite{Shuvaev}.
With the specific gluon distribution used by us, the prescription of \cite{Shuvaev} 
can be emulated by taking the ordinary gluon distribution at \cite{INS06}
\begin{equation}
x_{eff} = C_{skewed} \frac{M_V^2}{W^2}  \sim 0.41 \, \cdot \, {M_V^2 \over W^2} \,  .
\label{longitudinal_momentum_fraction}
\end{equation}
The full amplitude, at finite momentum transfer, 
well within the diffraction cone, 
is finally written as
\begin{eqnarray}
{\cal M}(W,\Delta^2) = (i + \rho) \, \Im m {\cal M}(W,\Delta^2=0) \, \exp(-B(W) \Delta^2) \, .
\end{eqnarray}
Here $\Delta^2$ is the (transverse) momentum transfer squared, $B(W)$ is
the energy--dependent slope parameter:
\begin{equation}
B(W) = B_0 + 2 \alpha'_{eff} \log \Big( {W^2 \over W^2_0} \Big) \, ,
\end{equation}
with 
$\alpha'_{eff} = 0.164$ GeV$^{-2}$ \cite{H1_Jpsi}, $W_0 = 95$ GeV. 
For the value of $B_0$ see the
discussion of the numerical results below.
For the small size $b\bar b$ dipoles relevant to our problem, a fast rise
of the cross section can be anticipated, and it is important 
to include the real part, which we do by means of the 
analyticity relation
\begin{eqnarray}
\rho = {\Re e {\cal M} \over \Im m {\cal M}} =  
\tan \Big[ {\pi \over 2} \, { \partial \log \Big( \Im m {\cal M}/W^2 \Big) \over \partial \log W^2 } 
\Big]
= \tan \Big( {\pi \over 2 } \, \Delta_{\Pom} \Big)
\, .
\end{eqnarray}

Finally, our amplitude is normalized such, that the differential cross section
for $\gamma p \to V p$ is
\begin{eqnarray}
{ d \sigma(\gamma p \to V p) \over d \Delta^2} = {1 + \rho^2 \over 16 \pi} \, 
\Big| \Im m { {\cal M} (W,\Delta^2) \over W^2 } \Big|^2 \exp(-B(W) \Delta^2) \, ,
\end{eqnarray}
and thus 
\begin{equation}
\sigma_{tot}(\gamma p \to V p) = {1 + \rho^2 \over 16 \pi B(W)} \, 
\Big| \Im m { {\cal M}(W,\Delta^2) \over W^2 } \Big|^2
\, .
\end{equation}
\subsection{$b \bar b$ wave function of the $\Upsilon$ meson}
We treat the $\Upsilon, \Upsilon'$ mesons as $b\bar{b}$ $s$--wave states,
the relevant formalism of light--cone wavefunctions is reviewed in \cite{INS06}.
The vertex for the $\Upsilon \to b \bar b$ transition is taken as
\begin{eqnarray}
\varepsilon_\mu \, \bar u(p_b) \Gamma^\mu v(p_{\bar b}) = [M^2 - M_V^2] \, \psi_V(z,k^2) \, 
\bar u(p_b) \Big( \gamma^\mu - {p_b^\mu - p_{\bar b}^\mu \over M + 2 m_b} \Big)
v(p_{\bar b}) \, \varepsilon_\mu \, ,
\end{eqnarray}
where $\varepsilon_\mu$ is the polarization vector of the vector meson
$V = \Upsilon,\Upsilon'$.
and $p_{b, \bar b}^\mu$ are the on-shell four--momenta of the $b,\bar b$ quarks, 
$p_{b, \bar b}^2 = m_b^2$.
The so--defined radial wave--function $\psi_V(z,k^2)$ can be regarded as a function
not of $z$ and $k^2$ independently, but rather of the three--momentum $\vec{p}$
of, say, the quark in the rest frame of the $b \bar b$ system of invariant mass $M$,
$ \vec{p} = (\bk , (2z-1) M/2) $.
Then,
\begin{eqnarray}
\psi_V(z,k^2) \to \psi_V(p^2) \, , \, {dz d^2 \bk \over z(1-z)} \to {4 \, d^3\vec{p} \over M} 
\, , p^2 = {M^2 - 4 m_b^2 \over 4} \, .
\end{eqnarray}
We assume that the Fock--space components of the $\Upsilon, \Upsilon'$--states
are exhausted by the $b \bar b$ components and impose on the light--cone 
wave function (LCWF) the orthonormality conditions ($i,j = \Upsilon,\Upsilon'$):
\begin{eqnarray}
\delta_{i j} = N_c \int {d^3 \vec p \over (2 \pi)^3} \, 4M \, \psi_i(p^2) \psi_j(p^2) \, . 
\end{eqnarray}
Important constraints on the LCWF are imposed by the decay width $V \to e^+ e^-$:
\begin{eqnarray}
\Gamma (V \to e^+ e^-) = {4 \pi \alpha_{em}^2 c^2_\Upsilon \over 3 M_V^3} \, \cdot g_V^2 \cdot K_{NLO} \, ,
\, \, K_{NLO} = 1 - {16 \over 3 \pi } \alpha_S(m_b^2) \, , 
\end{eqnarray}
where (\cite{INS06,Igor})
\begin{eqnarray}
g_V = {8 N_c \over 3} \int {d^3 \vec p \over (2 \pi)^3} (M + m_b) \, \psi_V(p^2) \, .
\end{eqnarray}
For the -- fully nonperturbative -- LCWF we shall try two different scenarios, 
following again the suggestions in \cite{INS06,Igor}.
Firstly, the Gaussian, harmonic--oscillator--like wave functions:
\begin{equation}
\psi_{1S}(p^2) = C_1 
\exp\left( - \frac{p^2 a_1^2}{2} \right) \, , \, 
\psi_{2S}(p^2) = C_2 (\xi_0 - p^2 a_2^2) 
\exp\left( - \frac{p^2 a_2^2}{2} \right) \, , 
\label{harmonic_oscillator_WF}
\end{equation}
and secondly, the Coulomb--like wave functions, with a slowly decaying
power--like tail:
\begin{equation}
\psi_{1S}(p^2) = {C_1 \over \sqrt{M}} \, {1 \over (1 + a_1^2 p^2)^2} \, , \, 
\psi_{2S}(p^2) = {C_2 \over \sqrt{M}} \, {\xi_0 - a_2^2 p^2 \over (1 + a_2^2 p^2)^3} \, .
\end{equation}
The parameters $a_i^2$ are obtained from fitting the decay widths
into $e^+ e^-$, whereas $\xi_0$, and therefore
the position of the node of the $2S$ wave function, is obtained
from the orthogonality of the $2S$ and $1S$ states.
We used the following values for masses and widthes:
$M(\Upsilon(1S))= 9.46$ GeV, $M(\Upsilon(2S)) = 10.023$ GeV, $\Gamma(\Upsilon(1S)
\to e^+ e^- ) = 1.34$ keV, $\Gamma(\Upsilon(2S) \to e^+ e^-) = 0.61$ keV
\cite{PDG}.
\subsection{Numerical results and comparison with HERA data}
\label{Sec:results_HERA}
\begin{figure}[!h]  
\includegraphics[width=0.45\textwidth]{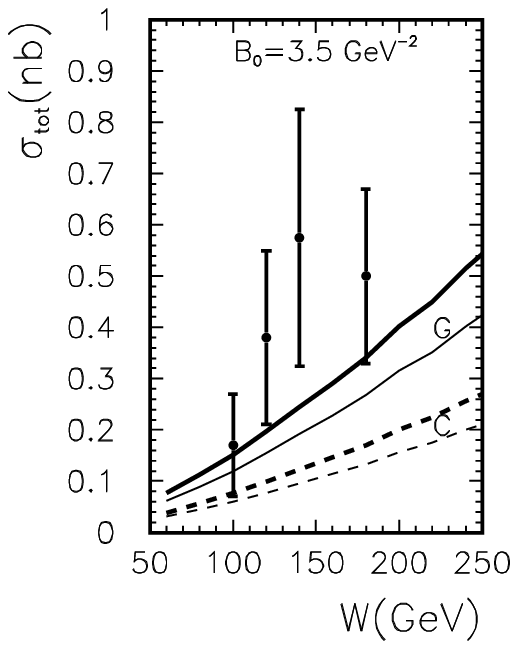}
\includegraphics[width=0.45\textwidth]{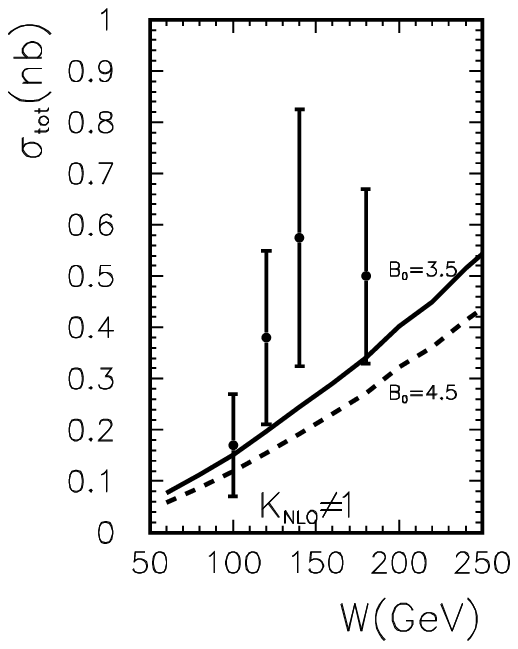}
   \caption{\label{fig:sig_tot}
   \small $\sigma_{tot} (\gamma p \to \Upsilon(1S) p)$ as a function
   of the $\gamma p$ cm--energy versus HERA--data. Left: dependence 
on the treatment of the $b \bar b \to \Upsilon$ transition; solid curves:
Gaussian (G) wave function, dashed curves: Coulomb--like (C) wave function.
Thick lines were obtained including the NLO--correction for the $\Upsilon$ decay width, while
for the thin lines $K_{NLO}=1$. Right: dependence on the 
slope parameter $B_0$ (given in GeV$^{-2})$, for the Gaussian wave function.
The experimental data are taken from \cite{ZEUS_old,H1,ZEUS_Upsilon}
}
\end{figure}
\begin{figure}[!h]  
\includegraphics[width=0.45\textwidth]{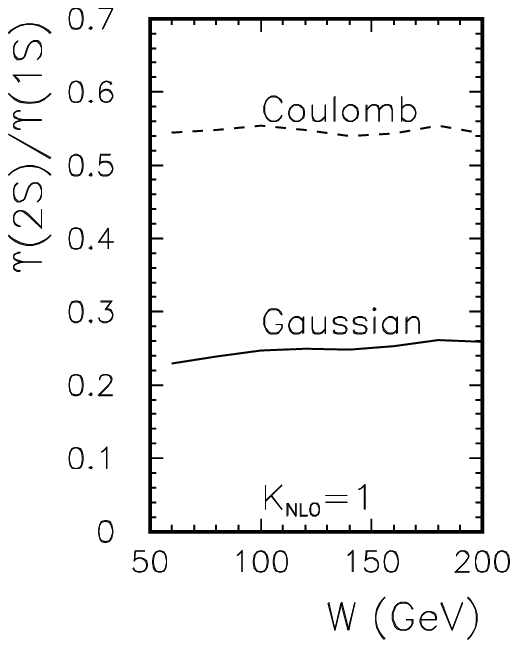}
\includegraphics[width=0.45\textwidth]{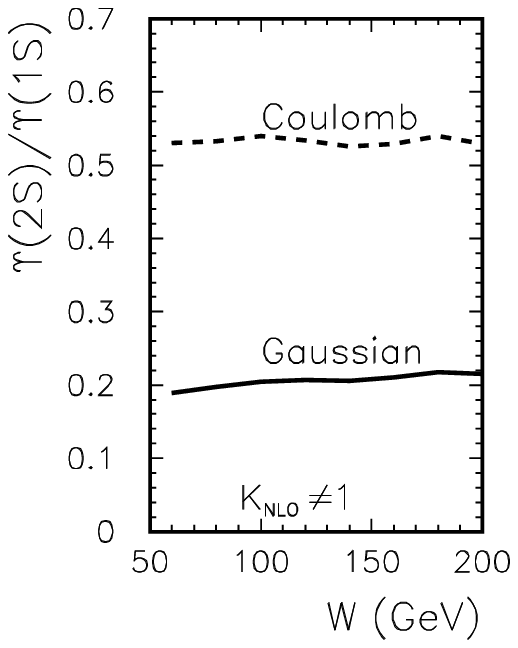}
   \caption{\label{fig:ratio_2S1S}
   \small The $2S/1S$-ratio $\sigma_{tot}(\gamma p \to \Upsilon(2S) p)/ 
\sigma_{tot}(\gamma p \to \Upsilon(1S) p )$ as a function
   of the $\gamma p$ cm--energy. }
\end{figure}
In Fig.\ref{fig:sig_tot} we show the total
cross section for the exclusive $\gamma p \to \Upsilon p$ 
process as a function of the $\gamma p$ cm-energy.
In the left panel we show results for two different 
wave functions discussed in the text: 
Gaussian (solid lines) and Coulomb-like (dashed lines).
Free parameters of the wave function have been
adjusted to reproduce the leptonic decay width in two
ways: (a) using leading order formula (thin lines)
and (b) inlcuding QCD corrections (thick lines).
Including the $K_{NLO}$--factor in the width 
enhances the momentum--space integral over the wave
function (the WF at the spatial origin), and hence
enhances the prediction for the photoproduction cross
section. Notice that strictly speaking inclusion of the
$\alpha_S$--correction is not really warranted given that
we do not have the corresponding radiative corrections
to the production amplitude.
Fortunately, due to the large scale $m_b^2$, the ambiguity 
in the two ways of adjusting the 
wave function parameters leads to only a marginal difference
in the total cross section over most of the relevant 
energy range.
To be fair, it should be mentioned, that the situation with 
the next--to--leading order corrections to diffractive
vector mesons is not a very comfortable one, see for example 
the instabilities reported in \cite{Ivanov_NLO}.  
But then, the systematic extension of $k_\perp$--factorisation
is yet lacking, so that at present we must be content with
estimates of the theoretical uncertainties obtained by changing
the principal parameters in the calculation.

As can bee seen from the figure, different functional
forms of the LCWF can lead to a quite substantial differences
in the predicted cross section. 
Finally, the absence of experimental data for $t$--distributions 
leaves the slope parameter $B_0$ only badly constrained. 
The full, energy dependent slope can be decomposed into three
contributions: one from the transition $\gamma \to V$,
a second one from the dynamics of the gluon ladder exchanged --
which induces the main part of its energy dependence, and
a third one from the elastic $p \to p$ vertex.
In comparison to $J/\psi$--production, we may expect, that
the slope in our case receives a smaller contribution from
the $\gamma \to V$ transition, due to the smaller transverse 
sizes involved \cite{NNPZZ}. 
It may therefore be expected that $B_0$ should be somewhat
smaller than in $J/\Psi$ photoproduction, where it is $\sim 4.6$
GeV$^{-2}$ \cite{H1_Jpsi}.  

We show the sensitivity
to the slope parameter $B_0$ in the right panel 
of Fig.\ref{fig:sig_tot}. 

We observe, that in general our predictions are systematically
somewhat below the experimental data. In principle, the agreement
could be improved by choosing an abnormally small value
for $B_0$, we shall however refrain from such an option.
In our view the description of data, given the large error bars,
is quite acceptable. The energy dependence of our result
corresponds to an effective $\Delta_\Pom \sim 0.39$.
For our predictions for Tevatron we shall use the Gaussian LCWF 
option, with the NLO correction to the width included.

In Fig.\ref{fig:ratio_2S1S} we show the ratio
of the cross section for the first radial excitation $\Upsilon(2S)$
to the cross section for the ground state $\Upsilon(1S)$.
The principal reason behind the suppression of the
$2S$ state is the well--known node effect (see \cite{NNPZ} 
and references therein) -- a
cancellation of strength in the $2S$ case due to the 
change of sign of the radial wave function.
It is perhaps not surprising, that the numerical
value of the $2S/1S$--ratio is strongly sensitive to the
shape of the radial light--cone wave function.

Here we assumed an equality of the slopes 
for $\Upsilon(1S)$ and $\Upsilon(2S)$ production. 
This appears to be justified, given 
the large spread of predictions from different wave functions.
We finally note, that the ratio depends very little on the 
choice of the $K_{NLO}$ factor (compare left and right panel). 
\section{Exclusive photoproduction in $p \bar p$ collisions}
\subsection{The absorbed $2 \to 3$ amplitude}
\label{2_to_3}
\begin{figure}[!h]    %
\includegraphics[width=0.8\textwidth]{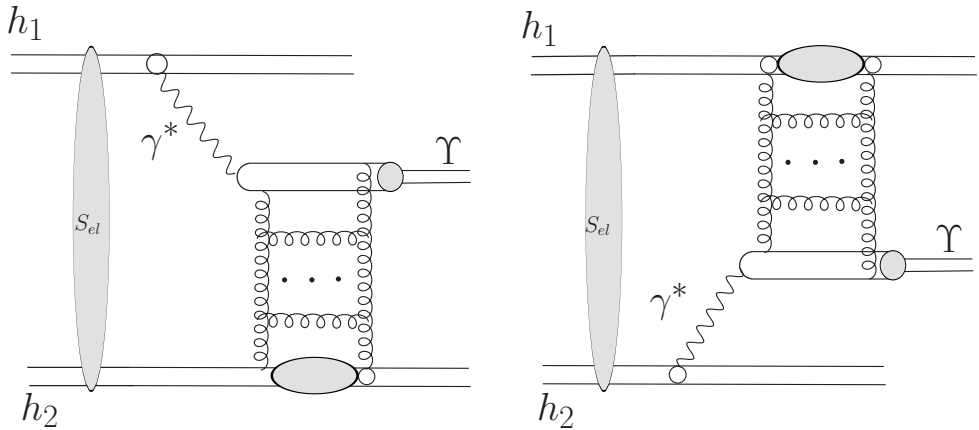}
   \caption{\label{fig:diagram_2}
   \small  A sketch of the two mechanisms considered in the present paper:
photon-pomeron (left) and pomeron-photon (right), including absorptive 
corrections.}
\end{figure}

The necessary formalism for the calculation
of amplitudes and cross--sections was outlined in sufficient detail
in Ref. \cite{SS07}. Here we give only a brief summary. 
The basic mechanisms are shown in Fig.\ref{fig:diagram_2}. 
The major difference from HERA, where the photon was emitted by a lepton
which does not participate in the strong interactions, now, both initial
state hadrons can be the source of the photon.  Therefore, it is now necessary
 to take account of the interference between two amplitudes. 
The photon exchange parts of the amplitude, involve only very small,
predominantly transverse momentum transfers. 
In fact, here we concentrate on the kinematic domain, where the outgoing 
protons lose only tiny fractions $z_1,z_2 \ll 1$ of their longitudinal 
momenta, in practice $z \lsim 0.1$ means $y \lsim 3$.
In terms of the transverse momenta of outgoing hadrons,
$\bp_{1,2}$, the relevant four--momentum transfers are
$t_i = - (\bp_i^2 + z_i^2 m_p^2)/(1-z_i) \, , i = 1,2$,
and $s_1 \approx (1 -z_2) s$ and $s_2 \approx (1-z_1) s$ are the
familiar Mandelstam variables for the appropriate subsystems. 
Photon virtualities $Q_i^2$ are small (what counts here is that
$Q_i^2 \ll M_\Upsilon^2$), so that the contribution from 
longitudinal photons can be safely
neglected. Also, as mentioned above, we assume 
the $s$--channel--helicity conservation in the 
$\gamma^* \to \Upsilon$ transition.
In summary we present the $2 \to 3$ Born-amplitude (without absorptive corrections)
in the form of a two--dimensional vector 
(corresponding to the two transverse (linear)
polarizations of the final state vector meson):
\begin{eqnarray} 
\bM^{(0)}(\bp_1,\bp_2) &&= e_1 {2 \over z_1} {\bp_1 \over t_1} 
{\cal{F}}_{\lambda_1' \lambda_1}(\bp_1,t_1)
{\cal {M}}_{\gamma^* h_2 \to V h_2}(s_2,t_2,Q_1^2)   
+ ( 1 \leftrightarrow 2 )
\end{eqnarray}
Inclusion of absorptive corrections (the 'elastic rescattering')
leads in momentum space to the full, absorbed amplitude
\begin{eqnarray}
\bM(\bp_1,\bp_2) &&= \int{d^2 \bk \over (2 \pi)^2} \, S_{el}(\bk) \,
\bM^{(0)}(\bp_1 - \bk, \bp_2 + \bk)  
= \bM^{(0)}(\bp_1,\bp_2) - \delta \bM(\bp_1,\bp_2) \, .
\nonumber \\
\label{rescattering term}
\end{eqnarray}
With 
\begin{equation}
S_{el}(\bk) = (2 \pi)^2 \delta^{(2)}(\bk) - \half T(\bk) \, \, \, ,
\, \, \, T(\bk) = \sigma^{p \bar p}_{tot}(s) \, \exp\Big(-\half B_{el} \bk^2 \Big) \, ,
\end{equation}
where $\sigma^{p \bar p}_{tot}(s) = 76$ mb, $B_{el} = 17 $ GeV$^{-2}$
\cite{CDF} ,
the absorptive correction $\delta \bM$ reads 
\begin{eqnarray}
\delta \bM(\bp_1,\bp_2) = \int {d^2\bk \over 2 (2\pi)^2} \, T(\bk) \,
\bM^{(0)}(\bp_1-\bk,\bp_2+\bk) \, .
\label{absorptive_corr}
\end{eqnarray}
The differential cross section is given in terms of $\bM$ as
\begin{equation}
d \sigma = { 1 \over 512 \pi^4 s^2 } | \bM |^2 \, dy dt_1 dt_2
d\phi \, ,
\end{equation}
where $y$ is the rapidity of the 
vector meson, and $\phi$ is the angle between $\bp_1$ and $\bp_2$.
\subsection{Results for Tevatron}
\begin{figure}[!h]  
\includegraphics[width=0.45\textwidth]{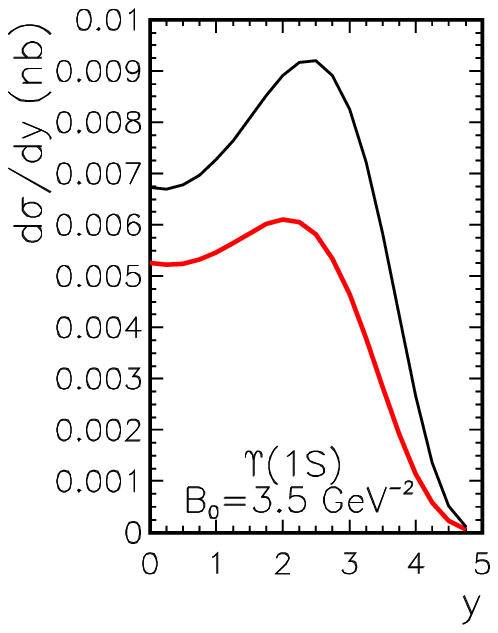}
\includegraphics[width=0.45\textwidth]{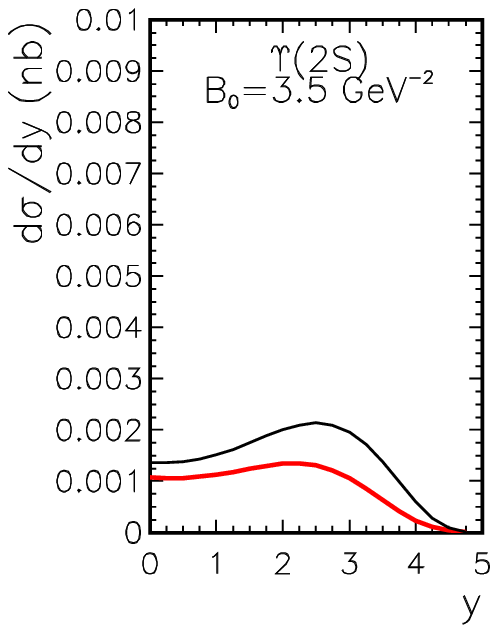}
   \caption{\label{fig:dsig_dy}
   \small Differential cross section $d\sigma / dy$ for $\Upsilon(1S)$ (left panel)
   and $\Upsilon(2S)$ (right panel) for the Tevatron energy
   $W$ = 1960 GeV. The thin solid line is for the calculation
  with bare amplitude, the thick line for the calculation with
  absorption effects included.}
\end{figure}
\begin{figure}[!h]  
\includegraphics[width=0.45\textwidth]{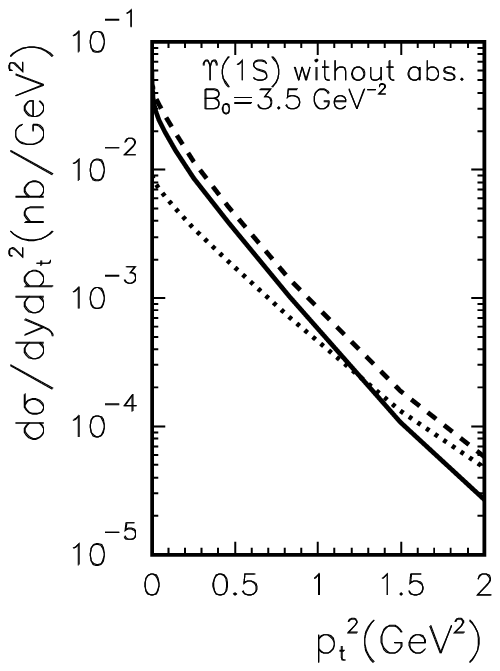}
\includegraphics[width=0.45\textwidth]{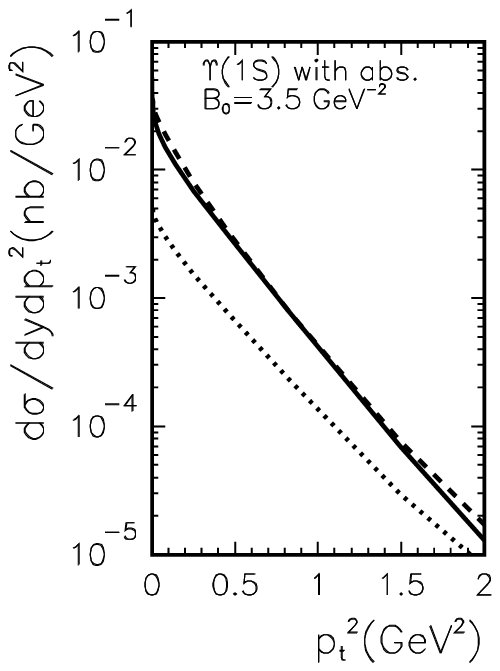}
   \caption{\label{fig:dsig_dpt2_U1S}
   \small
 Invariant cross section $d \sigma /dy dp_t^2$ for as a function of
$p_t^2$ for $\Upsilon(1S)$ for $W$ = 1960 GeV.
 The solid line: $y=0$, dashed line: $y=2$, dotted line: $y=4$.
Left panel: without absorptive corrections; Right panel: with 
absorptive corrections.
}
\end{figure}
\begin{figure}[!h]  
\includegraphics[width=0.45\textwidth]{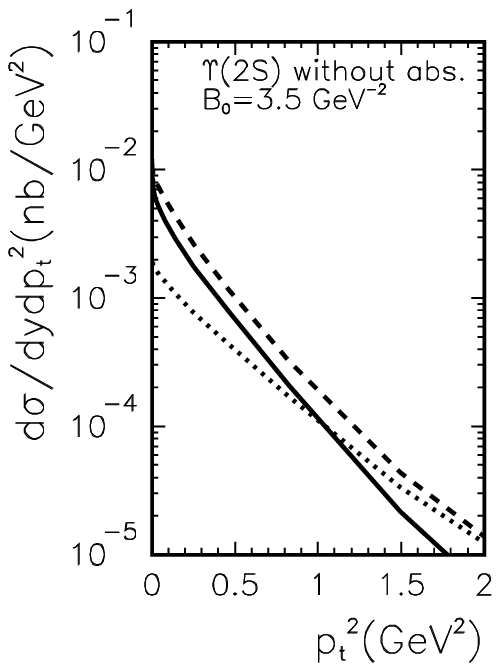}
\includegraphics[width=0.45\textwidth]{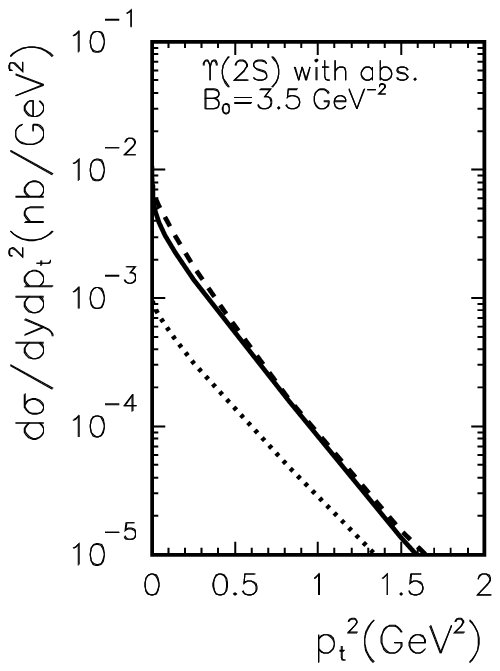}
   \caption{\label{fig:dsig_dpt2_U2S}
   \small
 Invariant cross section $d \sigma /dy dp_t^2$ for as a function of
$p_t^2$ for $\Upsilon(2S)$ for $W$ = 1960 GeV.
 The solid line: $y=0$, dashed line: $y=2$, dotted line: $y=4$.
Left panel: without absorptive corrections; Right panel: with 
absorptive corrections.
}
\end{figure}
\begin{figure}[!h]  
\includegraphics[width=0.45\textwidth]{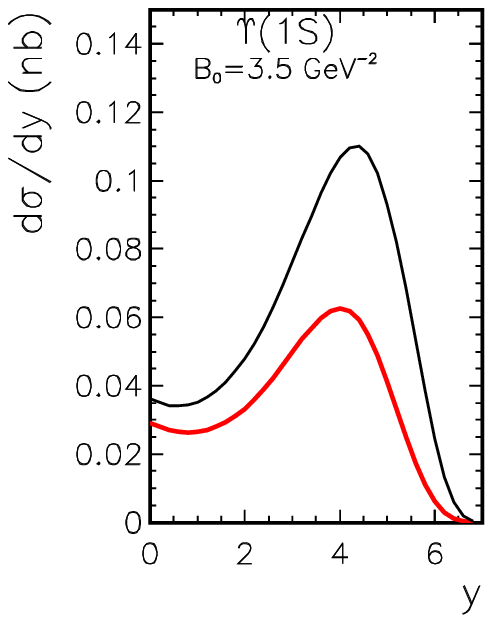}
\includegraphics[width=0.45\textwidth]{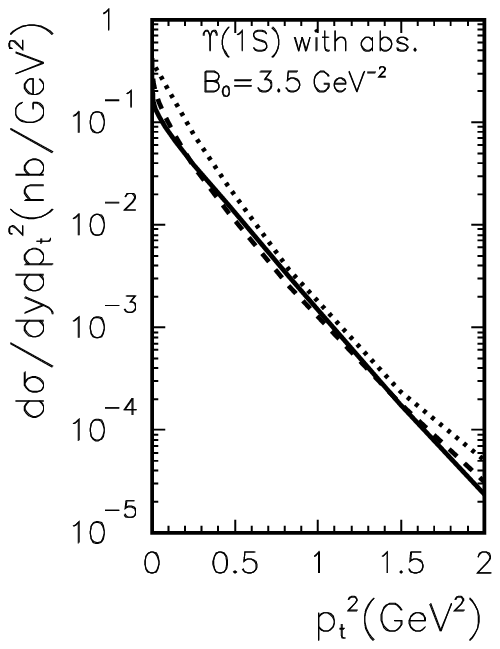}
   \caption{\label{fig:dsig_LHC}
   \small
Left panel: differential cross section $d \sigma / dy$ for $\Upsilon (1S)$
for the LHC energy $W= 14$ TeV. The thin solid line: without absorptive corrections;
thick line: with absorptive corrections. Right panel:
invariant cross section $d \sigma /dy dp_t^2$ for $\Upsilon(1S)$ 
as a function of $p_t^2$ for $W = 14$ TeV.
 The solid line: $y=0$, dashed line: $y=2$, dotted line: $y=4$.
Absorptive corrections are included.
}
\end{figure}
We now come to the results of differential cross sections for 
$\Upsilon$ production.
In Fig.\ref{fig:dsig_dy} we show the distribution
in rapidity of $\Upsilon(1S)$ (left panel) and
$\Upsilon(2S)$ (right panel). The ratio between $2S$ and
$1S$ follows closely the photoproduction ratio discussed 
in Sec. \ref{Sec:results_HERA}. 
The parameters chosen
for this calculation correspond to the Gaussian wave function, 
with $K_{NLO}$ included in the adjustment to the decay width.
Also the unintegrated gluon distribution is the same as the one
used  in section \ref{Sec:results_HERA}. 
The results obtained with bare amplitudes are shown by 
the thin (red) lines, and the results with absorption 
effects included are shown by thick (black) lines. 
Here the absorption effects are truly
a correction and cause only about 20-30\% decrease of 
the cross section. 
This is in sharp contrast to the situation for the fusion 
of two QCD ladders (relevant for the production of scalar 
charmonia or Higgs boson).
The rapidity distribution is only slightly
distorted by absorptive corrections. Notice that larger 
rapidities mean also larger photon virtualities and therefore 
somewhat smaller transverse distances in 
the $p \bar p$ collision are relevant.

Finally, in the following figures we show
distributions of $\Upsilon$'s in transverse momentum.
We show results for different values of rapidity:
$y = 0$ (solid), $y = 2$ (dashed) and $y = 4$ (dotted).
In Fig.\ref{fig:dsig_dpt2_U1S} we show the distributions
for $\Upsilon(1S)$ and in Fig.\ref{fig:dsig_dpt2_U2S}
for $\Upsilon(2S)$. Both, results with bare amplitudes
(left panels), and with absorption (right panels)
are shown. The inspection of the figures shows
that absorption effects are larger for large values
of the $\Upsilon$ transverse momenta -- they can 
lower the cross section by almost an order of magnitude at the largest
transverse momenta.
There is again a different effect of absorption for different 
rapidities. 

Notice, that our predictions, which use the 
low--$z$ approximation of the photon flux are most 
accurate at $y \lsim 3$. This is quite appropriate
for Tevatron, where it seems that a measurement is possible 
only at rather low rapidities.
We do not show here observables related to outgoing
proton or/and antiproton as they cannot be studied
experimentally at the Tevatron.
There will be, however, such a possibility at the LHC.

There are important issues regarding the extrapolation to 
LHC energies. Firstly the energy of the $ \gamma p \to \Upsilon p$ process
can vastly exceed the HERA range, and secondly the much increased
rapidity range may increase the importance of high--mass
diffraction for the absorptive corrections. Still, to give the reader
a rough idea of the expected cross section, we show
in Fig. \ref{fig:dsig_LHC} selected spectra at the LHC energy of $W = 14$ TeV.
Here, in the absorptive corrections, we used a Pomeron intercept
of $\Delta_\Pom = 0.08$.
It is interesting to point out that the
rise towards the maximum in the rapidity dstribution reflects the 
energy dependence of the $ \gamma p \to \Upsilon p$ subprocess.
Absorptive corrections in that subprocess, which we neglected so far 
can possibly alter the shape of the rapidity distribution. 
Since there are many other interesting aspects at
larger energies we leave a more detailed analysis for LHC for 
a separate publication. 

A brief comment on previous works is in order.
In \cite{Klein,Bzdak,GM} absorptive corrections were not included. 
The equivalent photon approximation is used in \cite{Klein,GM},
which allows only to obtain rapidity spectra. The form of the transverse
momentum distribution suggested in \cite{Klein} is not borne out by our
calculation.
Cross sections $d\sigma/dy$ obtained in \cite{Klein,Bzdak,GM} 
lie in the same ballpark as the results presented here.
However the shape of the rapidity distribution in \cite{GM}
is different from ours. 
\section{Conclusions}
We have calculated the forward amplitude for $\gamma p \to \Upsilon p$
reaction within the formalism of $k_\perp$-factorization. In this approach
the energy dependence of the process is encoded in the $x$-dependence
of unintegrated gluon distributions. The latter object is constrained by data 
on inclusive deep inelastic scattering.
The $t$-dependence for the $\gamma p \to \Upsilon p$ process involves
a free parameter and is in effect parametrized. 
Regarding the $\gamma \to \Upsilon$ transition, we used
different Ans\"atze for the $b \bar b$ wave functions.
The results for $\Upsilon(1S)$ production depend only slightly on the 
model of the wave function, while the $2S/1S$ ratio shows a substantial
sensitivity. We compared our results for the total cross section with a recent
data from HERA. Our results are systematically somewhat lower than data,
although the overall discrepancy is not worrysome, given the large uncertainties
due to the rather poor experimental resolution in the meson mass. 
The amplitudes for the $\gamma p \to \Upsilon p$ process are used
next to calculate the amplitude for the $p \bar p \to p \bar p \Upsilon$
reaction assuming the photon-Pomeron (Pomeron-photon) underlying dynamics.
In the present approach the Pomeron is then described within QCD in terms
of unintegrated gluon distributions.
We have calculated several differential distributions including soft 
absorption effects not included so far in the literature.
Our predictions are relevant for current experiments at the Tevatron,
predictions were made -- with qualifications -- for possible future experiments
at the LHC.

\section{Acknowledgements}
This work was partially supported by the Polish Ministry 
of Science and Higher Education (MNiSW) under contract 
1916/B/H03/2008/34.




\begin{thebibliography}{100}

\bibitem{quarkonia_reviews}
M. Kr\"amer, Prog. Part. Nucl. Phys. {\bf 47} (2001) 141;
J.P. Lansberg, Int. J. Mod. Phys. {\bf A21} (2006) 3857.

\bibitem{SMN}
  A.~Sch\"afer, L.~Mankiewicz and O.~Nachtmann,
  Phys.\ Lett.\  B {\bf 272}, 419 (1991).

\bibitem{KMR}
  V.~A.~Khoze, A.~D.~Martin and M.~G.~Ryskin,
  Eur.\ Phys.\ J.\  C {\bf 24}, 459 (2002).

\bibitem{Klein}
  S.~R.~Klein and J.~Nystrand,
  Phys.\ Rev.\ Lett.\  {\bf 92}, 142003 (2004).

\bibitem{GM05}
  V.~P.~Goncalves and M.~V.~T.~Machado,
  Eur.\ Phys.\ J.\  C {\bf 40}, 519 (2005).

\bibitem{Bzdak}
  A.~Bzdak, L.~Motyka, L.~Szymanowski and J.~R.~Cudell,
  Phys.\ Rev.\  D {\bf 75}, 094023 (2007).

\bibitem{SS07}
  W.~Sch\"afer and A.~Szczurek,
  Phys.\ Rev.\  D {\bf 76}, 094014 (2007).

\bibitem{GM}
  V.~P.~Goncalves and M.~V.~T.~Machado,
  Phys.\ Rev.\  D {\bf 77}, 014037 (2008).

\bibitem{ZEUS_old}
  J.~Breitweg {\it et al.}  [ZEUS Collaboration],
  Phys.\ Lett.\  B {\bf 437}, 432 (1998).

\bibitem{H1}
  C.~Adloff {\it et al.}  [H1 Collaboration],
  Phys.\ Lett.\  B {\bf 483}, 23 (2000).

\bibitem{ZEUS_Upsilon}
I. Rubinskiy for the H1 and ZEUS Collaborations, 
``Exclusive Processes in ep collision at HERA'', 
a talk at the International Europhysics Conference 
On High Energy Physics (EPS-HEP2007), Manchester, 
England, 19-25 July 2007.   

\bibitem{INS06}
 I.~P.~Ivanov, N.~N.~Nikolaev and A.~A.~Savin,
  Phys.\ Part.\ Nucl.\  {\bf 37}, 1 (2006).

\bibitem{IN_glue}
  I.~P.~Ivanov and N.~N.~Nikolaev,
  Phys.\ Rev.\  D {\bf 65}, 054004 (2002).

\bibitem{Unintegrated}
  J.~R.~Andersen {\it et al.}  [Small x Collaboration],
  Eur.\ Phys.\ J.\  C {\bf 48} (2006) 53;
 M.~\L uszczak and A.~Szczurek,
  Phys.\ Rev.\  D {\bf 73} (2006) 054028;
  A.~Szczurek,
  Acta Phys.\ Polon.\  B {\bf 34} (2003) 3191.

\bibitem{Pinfold}
 J.~Pinfold, a talk at Photon 2007: International Conference on the Structure 
and Interactions of the Photon and the 17th International Workshop on 
Photon-Photon Collisions and International Workshop on High Energy Photon Linear 
Colliders, Paris, France, 9-13 Jul 2007. 


\bibitem{H1_Jpsi}
  A.~Aktas {\it et al.}  [H1 Collaboration],
  Eur.\ Phys.\ J.\  C {\bf 46}, 585 (2006).

\bibitem{Shuvaev}
  A.~G.~Shuvaev, K.~J.~Golec-Biernat, A.~D.~Martin and M.~G.~Ryskin,
  Phys.\ Rev.\  D {\bf 60}, 014015 (1999).

\bibitem{Igor}
  I.~P.~Ivanov,
  ``Diffractive production of vector mesons in deep inelastic scattering
  within k(t)-factorization approach,''
  arXiv:hep-ph/0303053.

\bibitem{PDG}
  W.~M.~Yao {\it et al.}  [Particle Data Group],
  J.\ Phys.\ G {\bf 33}, 1 (2006).

\bibitem{Ivanov_NLO}
  D.~Y.~Ivanov, A.~Sch\"afer, L.~Szymanowski and G.~Krasnikov,
  Eur.\ Phys.\ J.\  C {\bf 34}, 297 (2004).

\bibitem{NNPZZ}
  J.~Nemchik, N.~N.~Nikolaev, E.~Predazzi, B.~G.~Zakharov and V.~R.~Zoller,
  J.\ Exp.\ Theor.\ Phys.\  {\bf 86}, 1054 (1998)
  [Zh.\ Eksp.\ Teor.\ Fiz.\  {\bf 113}, 1930 (1998)].

\bibitem{NNPZ}
  J.~Nemchik, N.~N.~Nikolaev, E.~Predazzi and B.~G.~Zakharov,
  Z.\ Phys.\  C {\bf 75}, 71 (1997)

\bibitem{CDF}
  F.~Abe {\it et al.}  [CDF Collaboration],
  Phys.\ Rev.\  D {\bf 50}, 5518 (1994).

\end{thebibliography}
\end{document}